# Off-Line, Multi-Detector Intensity Interferometers II: Implications and Applications

Aviv Ofir[*], Astronomy Dept, Tel Aviv University, Tel Aviv 69978 Israel,
and Erez N. Ribak, Physics Dept., Technion, Haifa 32000, Israel

ABSTRACT

Intensity interferometry removes the stringent requirements on mechanical precision and atmospheric corrections that plague all amplitude interferometry techniques at the cost of severely limited sensitivity. A new idea we recently introduced, very high redundancy, alleviates this problem. It enables the relatively simple construction (~1cm mechanical precision) of a ground-based astronomical facility able to transform a two-dimensional field of point-like sources to a three-dimensional distribution of micro-arcsec resolved systems, each imaged in several optical bands. Each system will also have its high resolution residual timing, high quality (inside each band) spectra and light curve, emergent flux, effective temperature, polarization effects and perhaps some thermodynamic properties, all directly measured. All the above attributes can be measured in a single observation run of such a dedicated facility. We conclude that after three decades of abandonment optical intensity interferometry deserves another review, also as a ground-based alternative to the science goals of space interferometers.

Subject Heading: instrumentation: interferometers - instrumentation: high angular resolution - techniques: interferometric

1. INTRODUCTION

Amplitude interferometry is today's mainstream technique for high angular resolution astronomy, mainly in the infrared regime. In principle, amplitude interferometers combine the electromagnetic waves from two or more separate locations to produce a high-resolution brightness distribution, or image, of the source. Intensity interferometry, on the other hand, combines the *intensities* of the electromagnetic wave via the correlation of the electrical currents generated by the detectors of the already-detected intensities. For astronomical purposes, the main advantage of intensity interferometry is its mechanical robustness: the required mechanical accuracy depends on the electrical bandwidth of the detectors, and not on the wavelength of the light. This opto-mechanical robustness also means that the atmosphere does not influence the performance of the instrument. The main disadvantages of intensity interferometry are its very low intrinsic sensitivity and the fact that the classical, two-detector intensity interferometer cannot reconstruct the phase of the complex degree of coherence, and thus can't be used to produce true images [1].

In astronomy, intensity interferometry was initially used in the radio part of the spectrum [2]. Hanbury Brown and Twiss demonstrated that it would also work at visible wavelengths [3,4] and subsequently the Narrabri Stellar Intensity Interferometer (NSII) was constructed and operated from 1965 to 1976. The NSII was the first instrument to successfully measure the angular diameter of main sequence stars, and a total of 32 angular diameters were measured by the NSII [1]. The low sensitivity implied a limiting magnitude of $m_b$=2.5 despite the fact that NSII used a pair of large reflectors, 30m$^2$ each. In essence, this low sensitivity is due to the fact that intensity correlation is a second-order effect.

The three-detector intensity interferometer, which correlates the intensities from three separate detectors (and so is a third order effect), was first proposed by Gamo [5] as a way to reconstruct the phase of the complex degree of coherence, and this technique was subsequently proved in the laboratory by Sato *et al* [6]. Since then, many other algorithms were proposed for the full reconstruction of the complex degree of coherence from amplitude-only measurements [7-12]. Classical two-detector intensity interferometry [1] was abandoned in the mid seventies due to its low sensitivity, and indeed higher-order correlations were never observed for astronomical sources.

---

[*] Address for correspondence: avivofir@wise.tau.ac.il

## 2. OFF-LINE, MULTI-DETECTOR INTENSITY INTERFEROMETERS

### 2.1. Off-line processing

In 1983 Fontana generalized intensity interferometry to $N$ detectors correlating all $N$ currents to form a single output [13]. We built on Fontana's work and examined in the first paper of this series [14] a linear array of many detectors with a uniform spacing $d$. We showed that this highly redundant configuration, when coupled with off-line processing, can be used to effectively increase the overall signal to noise ratio (SNR) of the instrument (or alternatively, its sensitivity). It appears that the $N$-detector intensity interferometer can operate in a regime where the SNR scales approximately *exponentially* with the product $NA\alpha n$, where $A$ is the area of each reflector, $\alpha$ is the detector's quantum efficiency and $n$ is the spectral photon flux density of the source. For example, we simulate a $0^{th}$ magnitude source observed with a 100-element array of dishes, each with $A = 7680m^2$ (~100m diameter) for one hour. The data, processed as an intensity interferometer array using NSII's technology (1960s technology), will have a SNR of more than $1.2 \cdot 10^6$. Indeed, most of this performance is due to the very large active area of the array, but this is also more than 190 times better than what one would expect if one used the previously known scaling laws alone [1]. We emphasize that each of the ~100m diameter reflectors is not a telescope but a crude light bucket. This 190-fold improvement is due to the high redundancy of the uniform linear array and the algorithm we proposed to take advantage of that, which is different from what was done in radio telescopes such as Westerbork [15]. This high redundancy can be divided to two types: translational symmetry, which does not decrease for dimmer sources and contributes a five-fold improvement, and higher order correlations, which are very dependant on the source's spectral photon flux density and contribute the rest of the improvement. Therefore, the earlier type of redundancy will still contribute to the performance of the instrument for faint sources, whereas the later type will not. If we define the limiting magnitude of the above array as the magnitude where we only get SNR of 3 after one hour of integration, then the limiting magnitude will be slightly more than the $10^{th}$ magnitude using the NSII-old technology, and about 14.4 magnitudes when conservative technological improvements are considered (off-the-shelf detectors and electronics, see [14]). We also introduced in [14] a "coronagraphic" effect of intensity interferometry which can be used to increase the dynamic range of the instrument even further, but this effect was not thoroughly explored and is not included in the above-mentioned computations.

The SNR also scales with the number of optical channels $p^{\frac{1}{2}}$ and electrical bandwidth $b_v^{\frac{1}{2}}$ of the detectors [1]. Observations of several targets in a sparse field can be made simultaneously by having a sparse two dimensional detector array mounted on each reflector (similar to atmospheric Čerenkov telescopes), with each point-like source illuminating a different detector where all the $j$ detectors onboard each of the $N$ reflectors point at the same source. The detectors will be further apart then the reflector's point spread function (which will be quite wide, given the crudeness of the reflector).

Recording and offline processing all of data will enable a very productive scientific facility that will measure other observables simultaneously, on top of all quantities derived from precise astrometry:
1) Information from all reflectors is gathered simultaneously. The entire correlation function ($N$-1 points) will be measured in a single observation run (on $p$ optical bands). The many (tens and up) baselines will create a dense Fourier plane coverage in a single run, without the need to fit a model to the data. Phases could be recovered by one of the already known and somewhat redundant algorithms [7-12]. Imaging will be possible by having sufficient ($u, v$) coverage (see also §2.2). The proposed instrument will create images with a *minimum* angular resolutions of 1 micro-arcsec (100 elements, each 100m in diameter, so the baseline ≥ 10km), assuming all the reflectors have no spacing between them, and so probably even significantly better. The ultimate angular resolution limit will probably be set by the longest available baseline, and not by technical difficulty of using it. This accuracy can be compared to GAIA's 10 micro-arcsec and SIM's 1 micro arcsec maximum resolutions.
2) Averaging individual reflectors over relatively long periods (seconds) will provide high quality photometry of each target on each of the $p$ optical bands on the same run ($N$ redundant times). This is actually a spectrum with a low, $p$ point, resolution but high quality (high SNR) due to the very large collecting area. Observing several targets simultaneously will enable differential photometry.
3) Simultaneous observations in several optical channels will also enable differential interferometry [16] to extend even further the dynamical range of the instrument.
4) High resolution spectra (within each optical band) of each source can be recovered ($N$ redundant times) by correlation spectroscopy [13].
5) Multi-detectors intensity interferometers have an independent capability to measure the distance to each source by searching for the maximum signal [13]. This measurement will also give results on the same run (many times:

from all subgroups {$m$} with |$m$|≥3). Assuming we have good knowledge of the distance between the detectors $d/c$ (where $c$ is the speed of light), the relative error of each of the single distance measurements is dominated by the relative lateral (in the plane of the sky) error, as can be seen from [13]. Now we can try to take advantage of the fact that we have many individual distance measurements from many subgroups: It is interesting to compare the statistical significance of 1 earth-orbit parallax measurement ($3 \cdot 10^8$ km baseline) and many 1km baseline measurements – which happens at $(3 \cdot 10^8)^2$ individual measurements, or after using all relevant subgroups of an array with just 58 elements. Theoretically, a 100-element array with a 1km typical baseline will measure distances with the same statistical significance as a single measurement with a baseline of 84 light-years. Unfortunately, this is too optimistic since this computation assumes we can detect the distance measurement signal from *each individual* subgroup - which is very difficult already at the $3^{rd}$ order correlation and quite impossible at the $50^{th}$ order correlations (as explained in [14], even for the ~100m diameter reflectors the *total* of all $3^{rd}$ order correlations is negligible for sources dimmer than magnitude 3, when using NSII technology).

6) The high sampling rate of an intensity interferometer, all the way to the GHz scale, will enable the use of the residual timing technique for appropriate sources (for example: pulsars, stellar scintillation, eclipsing objects).
7) NSII was already used to measure directly the emergent flux of the source, the source's effective temperature and effect of polarization [1]. These observables will also be available to the multi-detector intensity interferometer.
8) Multi-photon experiments may uncover thermodynamic information of how the light was originally created or how it has been scattered since its creation [17]. A multi-detector intensity interferometer may bring this type of information to the measurable range.
9) As a by product, an intensity interferometer can operate as an atmospheric Čerenkov camera. We comment that (a) as a Čerenkov detector, an intensity interferometer might be a bit out of focus as it will be focused to infinity, whereas Čerenkov detectors focus to an altitude of ~10km. (b) Čerenkov radiation will not contaminate long baselines (>300m).
10) When used for bright sources (not necessarily astronomical sources), this technique can by used as a probe to quantum optics via the pronounced multi-correlation (photon bunching) effect.

Building a completely non-redundant (or only partially redundant) array, in contrast to our fully-redundant array, will only affect the SNR of the reconstructed correlation function, and not the other observables. We also refer to other proposed algorithms related to intensity interferometry improvement:
- Improving the contrast between different parts of the image, and enlarging it without any limitation and, in spite of this enlargement, without any distortions of objects borders caused by discreteness of initial image [18].
- Implementation of an eigenfunction method to the problem of correlation function restoration from the photocurrent data [19].

Very advanced technology might be required for digitization, computation power and storage. The full analysis of an observation run by an intensity interferometer, as we propose, will be computer intensive. One way this problem may be addressed is with a distributed computing platform such as BOINC (Berkeley Open Infrastructure for Network Computing), which successfully runs several distributed computing projects, of which one is SETI@home [20].

## 2.2. ($u,v$) Coverage

Since imaging is one of the most important capabilities of any future interferometer, and since the long exposure times necessitate operating in "snapshot" mode (vs. Earth-scan mode), one must consider not just linear configurations but also *2D* configurations which cover large parts of ($u,v$) plane. Consider the configuration depicted in Figure 1: a double linear equidistant array with spacing $d$, whose intersection point is between two elements of one of the arms. This type of instrument should be capable of two types of configurations:

1) When the two arms, each with $N$ reflectors, form a single line, it is optimized for high SNR and dynamical range, acting as a $2N$ linear array, $d/2$ spacing.
2) When one arm is rotated at an angle with respect to the other: apart from two sets ($N$-1 points each) of high SNR points along each arm, one would get $N^2$ additional points in the ($u,v$) plane, equally spaced if the arms are perpendicular, from inter-arm baselines, but without redundancy for them. This configuration is optimized for geometrically complex objects that require better ($u,v$) coverage, for a total of $N^2 + 2(N-1)$ different ($u,v$) points, all observed simultaneously.

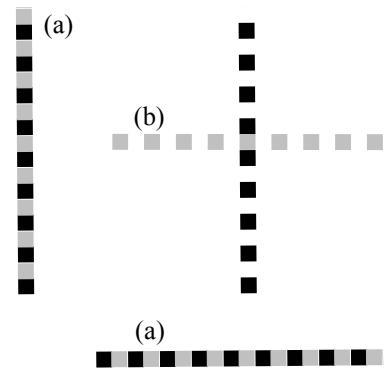

**Figure 1**: Two $N$-redundant linear arrays can be placed at any angle, here they are inline (a) and at $90^0$ (b). North is up.

For comparison, a completely non-redundant array with $2N$ reflectors would have $N(2N-1)$ different ($u,v$) points, less than twice the number of baselines as the double linear array, but will have no redundancy whatsoever (with its associated SNR benefits). We note that Herrero [21] suggested that a *V*-shaped array will be the optimal configuration for an intensity interferometer array.

2.3. Realization

It seams that in any case the reflectors will have to be mobile in order to keep the projected separations constant in spite of Earth's rotation, since fixed reflectors would force short integration periods and would render the instrument impotent. Indeed, for this very reason the two NSII reflectors moved on circular tracks during the observations [1]. Further simplification may come about by using any of these: mobile focal collector concept employed in radio astronomy, rotating liquid mirrors or the Carlina hypertelescope [22]. The crude reflectors adversely affect the ability to focus such large reflectors on the relatively small active area of the detectors. This disadvantage may be mitigated by using non-imaging concentrators [23], provided that the total path difference after these additional optical elements is not more than the specified tolerance (1cm in our example). Beyond these, we can imagine three major types of realizations: arrays with a central tower, arrays without a central tower, and a cross between the two:

1) Central tower, as in STACEE (Solar Tower Atmospheric Čerenkov Effect Experiment [24]): a field of "dumb" mirrors will simply reflect light to a central tower where all light will be detected and processed [1]. Advantages: Inexpensive, probably feasible by modifying or nearly replicating a solar tower experiment. The detectors are not mounted on the reflectors, so all the sensitive electronics of the entire instrument are fixed, protected and easily accessible, Almost flat reflectors, Shorter electrical leads from detection to recording. Disadvantages: Due to the poor optical quality of the reflectors and the large distance to the tower, probably only one target per observation will be possible, Atmospheric extinction of the signal and scintillation along the way from the mirrors to the tower.

2) No central tower, as in NSII: every reflector will carry all its detectors onboard. Advantages: The short reflector-detector distance enables multi-targeting, No atmospheric extinction or scintillation from the reflectors to detectors. Disadvantages: Cannot be realized on any existing array (including astronomical arrays observing at the mm or longer wavelengths), without major modifications, and probably would have to be a dedicated instrument, Will create a need to protect the detectors from the elements on many separate and mobile platforms.

3) A cross of the two may be possible: instead of placing a detector on every reflector on the NSII-like configuration, a light guide is positioned at the focus, for example, a multimode optical fiber or a bundle thereof, bringing the signals all the way to a central, fixed, lab. Advantages: it eliminates all the disadvantages of the two configurations above, except the need to build it probably as a dedicated facility. It might be relatively inexpensive. Disadvantages: the poor optical quality of the reflector may require non-imaging optics to collect the light into the light guide.

As a short-term goal, we think that the best place to start is STACEE by augmenting it with much more data storage capability (hours of real-time data, instead of the current ~100 nanoseconds). In practice, some work will also be done in order to eliminate any zero-point drifts of the detectors to a very high degree of accuracy. We think that STACEE is a prime candidate for early experiments to validate the results of this work since it already has a field of uniformly spaced (albeit fixed) reflectors, all the electronics of high-speed detection of light from specific reflectors, and of course, the fact that it is already build, so capital investment will be minimal indeed. Although the 37m$^2$ area of each of the 220 heliostats may seem too small, this limitation can be overcome by combining the detected intensity of several adjacent heliostats to simulate a single large reflector (after correcting for geometric path difference), and thus the whole field can simulate a few very large reflectors. This setup may also help to mitigate the focusing mismatch between atmospheric Čerenkov cameras and intensity interferometers. After this work was accomplished [25], a more limited proposal was also raised following the same lines [26].

Among many others, two special targets may utilize the capabilities of intensity interferometry to the fullest:
1) Fast repeaters (such as pulsars): as in boxcar averaging, binning the few-GHz of samples into millisecond bins, synchronized with the pulsar's period, will provide a way to create a "movie" of the different phases of the Pulsar. The cost: observing time will scale by the number of frames in the "movie".

2) Close binaries and multiples, including bright compact objects and extra-solar planets: these will be found and characterized by astrometry (from interferometry and imaging), radial velocity (from high resolution spectra), photometry (transits and lensing events), and residual timing, all independently and in the same observational run.

One of the appeals of the proposed instrument is that no new technology needs to be developed and all the components (besides the reflectors themselves) are off-the-shelf products. Yet, we can say what kind of technological advance would further promote this kind of instrument significantly: since the only uncontrolled parameter of the exponential factor $NA\alpha n$ is the spectral flux density $n$, then theoretically one would want to have a device that optically amplifies the intensity of the source before it is electronically detected, in front of every detector. The difficulty is that this device needs to operate at high speed, uniformly (for all detectors) and while keeping all the statistical properties of the light to allow for post-processing of the data. We note that amplification of an optical signal always involves its absorption first, and thus the above advantages can currently only be realized at much longer wavelengths ($\lambda \geq 30\mu m$), where off-line amplitude interferometers, with their superior sensitivity, might be feasible.

We note that such a large facility may have other uses during the daytime that may generate revenues and thus help to finance it's construction. These uses include uses usually associated with concentrating solar power plants, for example - power generation and garbage disposal by high temperature incineration.

3. CONCLUSIONS

We recently presented an algorithm for the improvement of the SNR of an evenly spaced off-line multi-detector intensity interferometer by utilizing its very high redundancy. We then showed that under the investigated conditions translational symmetry improves the performance of the proposed instrument by a factor of about five, and that multi-correlation can further improve that performance significantly, the total improvement reaching more than two orders of magnitude. This improvement is made possible by the offline processing of the data that allows us to utilise each photon several times and thus alleviate the low intrinsic sensitivity of intensity interferometers. We believe that this algorithm is not the only one possible. The intensity interferometer array proposed will record all information and process it offline, which will also allow the application of new algorithms to previous observations. Offline processing of the data will also enable measuring a variety of other observables, including photometry, spectroscopy, distance and timing, and thus to create a very productive astronomical facility.

Admittedly, the proposed instrument is big: 100 elements, each ~100m in diameter is not a simple thing to call for. Yet, it is very well within current technical capabilities as all of the requirements are already well-exceeded by different currently operating or already-planned instruments: STACEE is already using some of the 220 digitally-controlled $37m^2$ heliostats for optical detection at 1 GHz, The Green Bank Telescope is a single aperture that already boasts $7853m^2$ of collecting area, accurate to better than 0.22mm rms [27], compared to $7680m^2$ accurate to 1cm that we used in our calculations, and the total scale of the proposed instrument ($768,000m^2$ active area) is smaller than the proposed $10^6$ $m^2$ of the Square Kilometer Array at comparable mechanical accuracy [28]. We did not make any cost assessments for the construction of such a facility, but we recognize that cost will be a crucial issue for such a major facility.

Using all of these properties will enable the relatively simple construction of a ground based facility able to transform a 2*D* field of point-like sources to a 3*D* distribution of micro-arcsec resolved systems. Each of the systems will be truly imaged in *p* optical bands without a need to fit the visibility curve to some model, and it will also have its high quality spectra (inside each optical band), photometry, emergent flux, effective temperature, high resolution residual timing and polarization effects measured. All of these can be achieved in a single observation run of such a dedicated facility. The facility will not need adaptive optics, beam combiners, delay lines, precision optics and mechanics of almost any kind. In addition, due to their mechanical robustness intensity interferometers are far more amenable to use in shorter wavelength, and indeed NSII already operated at the blue band at 438.4nm.

The above-mentioned properties may warrant an evaluation of such a facility as an alternative to space interferometer missions (such as SIM [29]). The proposed instrument might outperform SIM in many parameters: it will have far better resolution, more observables, it will create images for all observed objects, it presents a much simpler technological challenge, and it will be ground-based.

Contemporary astronomy is plagued by the need to have optical surfaces smooth, and distances fixed to a fraction of the wavelength. Multi-detector optical intensity interferometry offers a way out of this restriction, even if not for the faintest of objects, offering:
- Ease of construction since mechanical accuracy depends on electrical bandwidth, and not on wavelength. Using mobile focal collectors may significantly reduce the moving mass of each reflector.
- All reflectors are identical and are not connected optically to the others. Furthermore, reflectors will probably be segmented, enabling "industrialized" parts manufacturing.
- Relatively easy to obtain very long baselines of many kilometers at 1cm mechanical accuracy.
- No new technological development is needed.

After 35 years, results obtained with intensity interferometry are still the state of the art, resolution-wise, and especially so in the blue, where amplitude interferometry is lacking. The main drawback of intensity interferometry is sensitivity, but using all of the proposed improvements and scaling laws improves the limiting magnitude from 2.5 at NSII to 14.4 of the proposed instrument. It seams that multi-detector intensity interferometry could be used as a present day technique answering present day questions, and indeed deserves another review.

4. Acknowledgment

This work is based upon a master's thesis by A. Ofir [25]. Parts of this work were supported by the European Interferometry Initiative through OPTICON (an EU Framework VI program). We would like to thank William Tango, the referee of the first paper of this series for his comments and suggestions, of which some were included in this paper. A. Ofir may be contacted by avivofir@wise.tau.ac.il.